\documentclass[aps,pre,reprint,groupedaddress,longbibliography]{revtex4-2}

\usepackage{graphicx} 
\usepackage{amsmath}

\begin{document}
	\newcommand{\K}{{\mathrm K}}
	\newcommand{\ud}{{\mathrm d}}
	\newcommand{\umod}{\mathrm{mod}}
	\newcommand{\sech}{\mathrm{sech}}


\title{System size synchronization}


\author{Mar\'{i}a Laura Olivera-Atencio}
\author{ Manuel Morillo}
\author{Jes\'us Casado-Pascual}
\email[]{jcasado@us.es}
\affiliation{F\'{\i}sica Te\'orica,Universidad de Sevilla, Apartado de Correos 1065, Sevilla 41080, Spain}


\date{\today}

\begin{abstract}
	In this work we bring out the existence of a novel kind of synchronization associated to the size of a complex system. A dichotomic
	random jump process associated to the dynamics of an externally driven stochastic system with $N$ coupled units is constructed. We define an output frequency and phase diffusion coefficient. System size synchronization occurs when the average output frequency is locked to the external one and the average phase diffusion coefficient shows a very deep minimum for a range of system sizes. Analytical and numerical procedures are introduced to study the phenomenon, and the results describe successfully the existence of system size synchronization. 
\end{abstract}


\maketitle

\section{Introduction}
The topic of synchronization has been amply studied from a variety of different perspectives because of its intrinsic theoretical interest, as well as its widespread applications~\cite{pikovsky:2001, schimanskygeier:2001, lindner:2004}. Understanding the synchronization present in systems helps to comprehend processes studied in fields linked to engineering, biology, or medicine, in addition to physics.  
For instance, synchronization plays a role in power-grid networks, as discussed in Ref.~\cite{motter:2013}.  In biological models, noise and interaction delays help to explain the spontaneous formation of clusters of synchronized spikings in homogeneous neuronal ensembles~\cite{franovic:2012} and in biological mobile phase oscillators~\cite{petrungaro:2019}. Hydrodynamic synchronization of spontaneously beating filaments with different waveforms ranging from sperm to cilia and \textit{Chlamydomonas} is discussed in~\cite{chakrabarti:2019}. In addition, synchronization seems to play a role in the description and control of some medical problems~\cite{budzinski:2019}.

The study of synchronization has been addressed for different regimes. For chaotic regimes, synchronization has been analyzed in Lorenz systems~\cite{cuomo:1993}, in a real set of synchronizing chaotic circuits~\cite{pecora:1990}, in coupled oscillators~\cite{valladares:2001} and lasers~\cite{jayaprasath:2018}, and more recently in sets of oscillators with imperfections~\cite{sugitani:2021}. The presence of noise may significantly interact with the synchronization mechanism~\cite{reguera:2002}. In particular,
certain systems are able to synchronize with external forces when the noise takes adequate values. This phenomenon, known as noise induced synchronization, has been analyzed in single particle systems~\cite{freund:2000,rozenfeld:2001,freundbarbay:2003,park:2004,casado1:2005,casado2:2005}, as well as in multiparticle systems~\cite{casadovazquez:2007}.  
The study of synchronization has also been extended to the quantum regime, being analyzed initially in driven systems in~\cite{goychuk:2006}. More recently, studies of quantum synchronization have been carried out in open quantum systems~\cite{karpat:2021} or via quantum machine learning~\cite{cardenas:2019}. Quantum phase synchronization has also been experimentally observed~\cite{laskar:2020}.

In the present work, we investigate a novel aspect of the synchronization phenomenon in stochastic complex system:  \textit{system size synchronization}. With this term, we refer to a system size dependent type of synchronization, between a multiparticle system and an external force. 
Specifically, here we will show that a set of classical coupled elements immersed in a highly noisy environment might synchronize with a rather weak external driving if the number of elements lies within a range of optimal system sizes.

\section{A model for system size synchronization}
We consider a model describing a
\emph{finite} set of $N$ interacting \emph{bistable} subsystems,
each of them characterized by a single degree of freedom $x_i$, with $i=1,\ldots,N$, whose dynamics is governed by the Langevin equations \cite{pikovsky:2002}
\begin{equation}
	\dot{x}_i=x_i-x_i^3+\frac{\epsilon}{N}\sum_{j=1}^N(x_j-x_i)+\xi_i(t)+F(t).
	\label{Eq1}
\end{equation}
Here, $\epsilon$ is the parameter defining the strength of the
interaction between subsystems, the $\xi_i(t)$'s are Gaussian white
noises with zero average and 
$\langle \xi_i(t)\xi_j(s)\rangle=2D\delta_{ij}\delta(t-s)$, $D$ being the noise
strength, and $F(t)=F(t+T)$ is an external driving force of period $T$
with a constant amplitude $A$ for the first half of a period and $-A$ for
the second one. The model considered in Ref.~\cite{pikovsky:2002} is similar to this one but with a sinusoidal driving force.

We focus our interest on the single global variable $X(t)=\sum_{j=1}^N x_j(t)/N$.
The nonlinearity of the dynamics prevents us to write an exact closed equation for $X(t)$. Nevertheless,  in the absence of external driving, the asymptotic behavior of the equilibrium probability distribution for the global variable $X$ in the limit $N\to \infty$ can be analyzed~\cite{desai:1978}. From that analysis, it follows that there are two regions in the $(D,\epsilon)$ space: one in which the system is in a disordered phase, with a steady-state value $X=0$, and a second one in which the system is in an ordered phase. In this last phase, there are three possible steady-state values of $X$, namely, $X=0$, which is unstable, and $X=\pm X_0$, with $X_0$ being a parameter that depends on $D$ and $\epsilon$. 

\begin{figure}[t]
	\centering{\includegraphics[scale=.42]{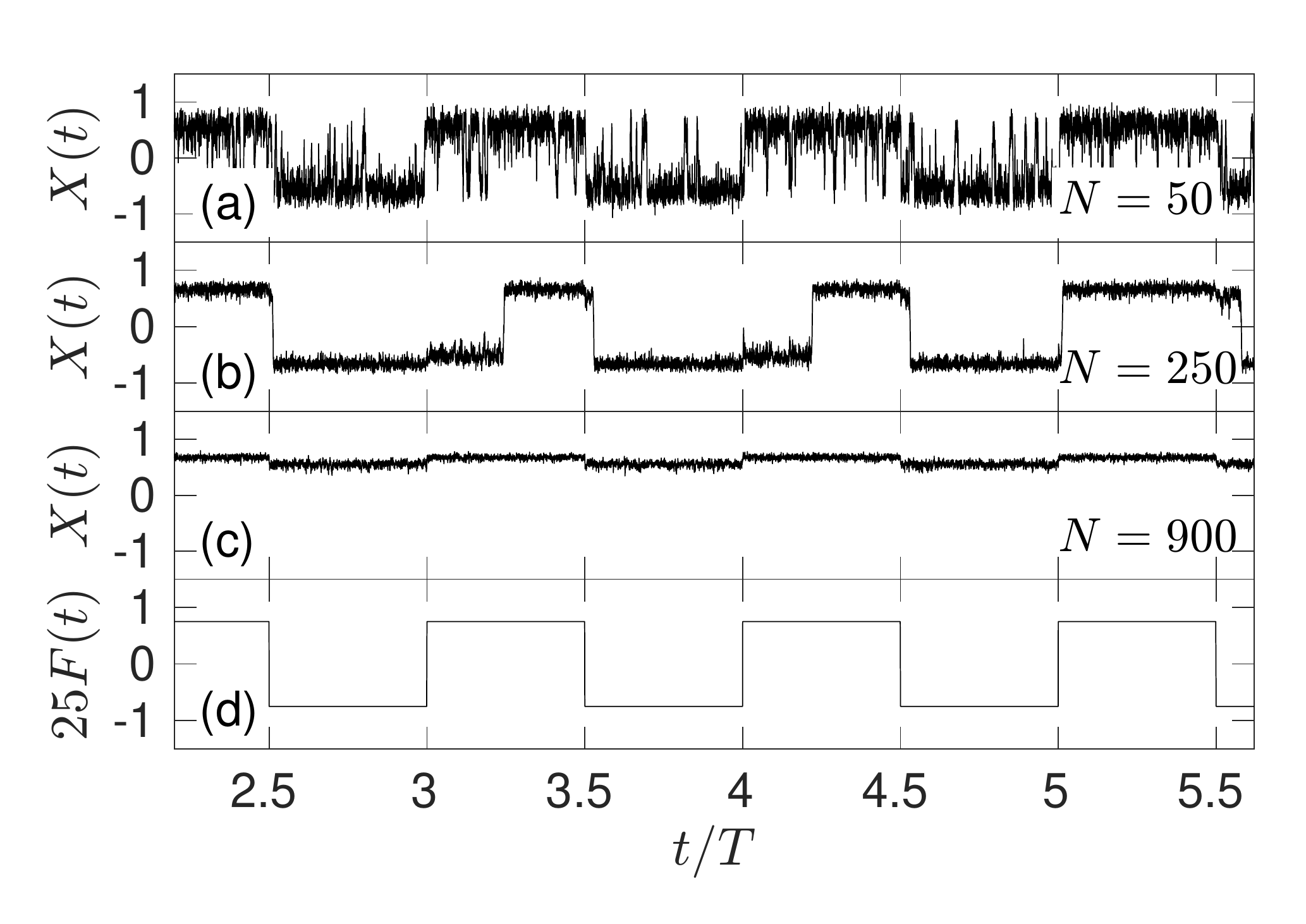}}
	\caption{Stochastic trajectories of the global variable $X(t)$ 
		for $N=50$ [panel~(a)], $N=250$ [panel~(b)], and $N=700$ [panel~(c)]. The remaining parameter values are: $\epsilon=2.7$, $D=0.7$, $A=0.03$, and $\omega=2\pi/T=0.001$.  Panel~(d) shows the external driving force conveniently amplified.}
	\label{Fig1}
\end{figure}

For finite system sizes, the above asymptotic results enable us to form a qualitative picture of the random behavior of $X(t)$.  In the ordered phase (the one we are interested in), $X(t)$ exhibits small fluctuations around the values $X_0$ and $-X_0$, and rather large fluctuations, or jumps, between these two values.  The rate of jumps decreases as $N$ increases and approaches zero in the limit $N\to \infty$. As shown in Fig.~\ref{Fig1}, this qualitative picture remains valid even in the presence of sufficiently weak external drivings.

In Ref.~\cite{pikovsky:2002}, Pikovsky et al. proposed a quantitative explanation of this last qualitative picture in terms of a standard noise-driven double-well model in which the noise strength is inversely proportional to the number of elements $N$. In addition, they showed that the response of $X(t)$ to the periodic external force exhibits a resonantlike behavior as $N$ is varied.  So, they demonstrated the existence of a phenomenon similar to stochastic resonance, but with the system size playing the role of the noise strength.

In this work, we are interested in the synchronization mechanism associated to the stochastic jump process. As discussed in Ref.~\cite{casado1:2005} for a single variable in a rocked, overdamped bistable potential, the stochastic synchronization phenomenon is quantified in terms of an output frequency and a phase diffusion coefficient. Specifically, the stochastic synchronization phenomenon is characterized by the existence of a range of noise strength values for which there is a matching of the output frequency and the driving frequency (noise-induced frequency locking), together with a sharp decrease of the phase diffusion coefficient (noise-induced phase locking). For many variables, the effective noise scaling with the inverse of $N$ prompts us to ask whether a phenomenon similar to stochastic synchronization may also show up as a function of the system size.

\begin{figure}[t]
	\includegraphics[scale=0.42]{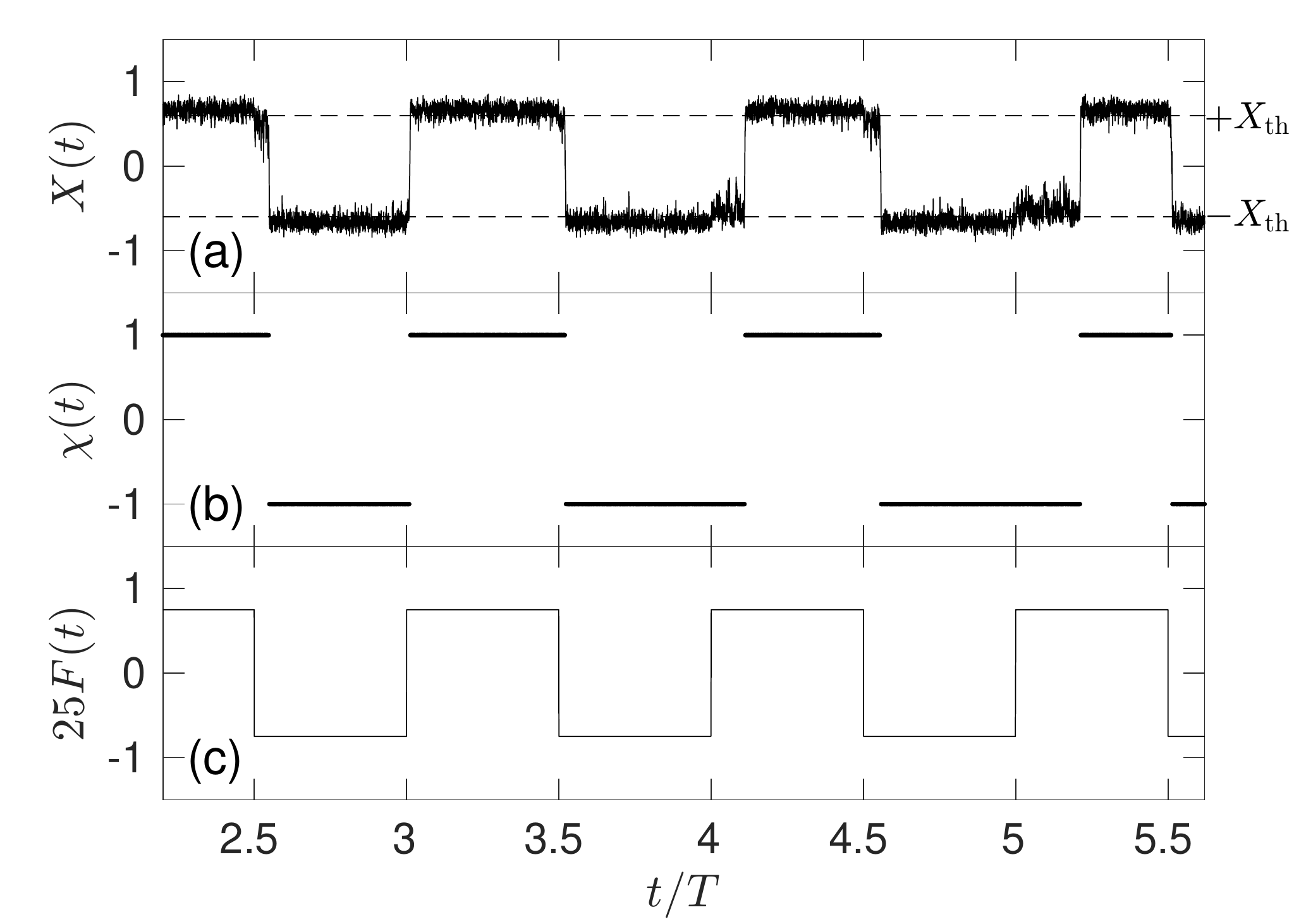}
	\caption{Sketch of the filtering process for a system with size $N=250$ and the same parameter values as in Fig.~\ref{Fig1}. Panel~(a) depicts a random trajectory and the corresponding threshold values $X_{\mathrm{th}}$ and $-X_{\mathrm{th}}$. The filtered trajectory $\chi(t)$ is depicted in panel~(b).  Panel~(c) shows the external driving force conveniently amplified.}
	\label{Fig2}
\end{figure}

To explore this possibility, we must adapt the definitions of output frequency and phase diffusion coefficient used in Ref.~\cite{casado1:2005} to many-variable systems. The first step is to introduce a discrete phase associated to the continuous stochastic process $X(t)$.  To this end, we proceed to filter out the small fluctuations of $X(t)$ to obtain a two-state stochastic process $\chi(t)$ taking the two values $+1$ or $-1$. The filtering process involves the consideration of two threshold values, $X_\mathrm{th}$ and  $-X_\mathrm{th}$, close to the levels of the small fluctuations (see Fig.~\ref{Fig2}). All the $N$ variables are initially located at $x_i(0)=X_\mathrm{th}$, so that $X(0)=X_\mathrm{th}$, and we assign $\chi(0)=+1$. A switch of $\chi(t)$ occurs whenever $X(t)$, having started in one of the threshold values, reaches the	other threshold value for the first time. The instant of time at which the $n$th switch takes place is a random variable which will be denoted by $\mathcal{T}_n$, with $n=1,2,\dots$. These random variables can be formally defined recursively as $\mathcal{T}_n=\mathrm{min}[t \,|\,  t>\mathcal{T}_{n-1}\;\mathrm{and}\;X(t)=(-1)^n X_\mathrm{th} ]$, with  $\mathcal{T}_0=0$. Next, we define a stochastic process $\mathcal{N}(t)$ by counting the number of switches within the interval $(0,t]$ as $\mathcal{N}(t)=\mathrm{max}(n \,|\, \mathcal{T}_n \le t)$. The filtered process is then given by $\chi(t)=\cos[\pi \mathcal{N}(t)]$. Associated to this filtered process, we define the stochastic phase $\varphi(t)=\pi \mathcal{N}(t)$,
the averaged output frequency
\begin{equation}
	\Omega_\mathrm{out}=\lim_{t\to+\infty}\frac{\left\langle \varphi(t)\right\rangle}{t},
	\label{freq} 
\end{equation}
and the averaged phase diffusion coefficient
\begin{equation}
	D_\mathrm{out}=\lim_{t\to+\infty}\frac{
		\langle \left[
		\varphi(t)\right]^2\rangle-\left\langle
		\varphi(t)\right\rangle^2}{t},
	\label{Diffcoeff} 
\end{equation}
where the angular brackets indicate averages over the random realizations. The system size synchronization refers to two features happening in a range of $N$ values: the matching of the  output frequency $	\Omega_\mathrm{out}$ to the driving one $\omega=2\pi/T$ and, simultaneously, very small values of the phase diffusion coefficient $D_\mathrm{out}$.

\section{Analytical results}
Assuming that $\chi(t)$ is a  Markovian dichotomic process, analytical expressions for $\Omega_\mathrm{out}$ and $D_\mathrm{out}$, completely determined by the rates of escape from both states, can be obtained~\cite{casado2:2005}. If we further assume that the rates of escape from the states $+1$ and $-1$ are, respectively, $\gamma_{+}$ and $\gamma_{-}$ for the first half of a period and $\gamma_{-}$ and $\gamma_{+}$ for the second one, it can be shown that~\cite{casado1:2005}
\begin{equation}
	\Omega_{\mathrm{out}}=\frac{\pi\gamma}{2}\left[1- \Delta^2 \left(1-\frac{\tanh\alpha}{\alpha}\right)\right]
	\label{anal_freq}
\end{equation}
and
\begin{multline}
	D_{\mathrm{out}} = \pi \Omega_{\mathrm{out}}- \frac{2\pi^2\Delta^4\tanh^3 \alpha}{T}-\frac{2\pi^2\Delta^2\left(1-\Delta^2\right)}{T}\\
	\times \left[3\tanh \alpha-\alpha\left( 1+2 \,\sech^2 \alpha  \right) \right],\label{anal_diff}
\end{multline}
where $\gamma=\gamma_{+}+\gamma_{-}$, $\Delta =(\gamma_{-}-\gamma_{+})/\gamma$, and $\alpha=\gamma T/4$.

To obtain analytical expressions for the rates of escape $\gamma_{+}$ and $\gamma_{-}$, the approximate analytical approach developed in Ref.~\cite{pikovsky:2002} will be used. In that reference, the authors derive a Langevin closed equation for $X(t)$ by applying a Gaussian approximation to the entire dynamics given by Eq.~(\ref{Eq1}), as well  as the slaving principle~\cite{haken:1982,haken:1996}. The resulting equation is
\begin{equation}
	\dot{X}(t)=aX(t)-bX^3(t)+\eta(t)+F(t),
	\label{Lang1v}
\end{equation}  
where $\smash{a=(\epsilon+1-c)/2}$, $\smash{b=[-1+3(\epsilon-1)/c]/2}$, with $\smash{c=\sqrt{(\epsilon-1)^2+12D}}$, and $\eta(t)$ is a Gaussian white noise with zero mean and $\langle \eta(t) \eta(t') \rangle=2D\delta(t-t')/N$. For the derivation of Eq.~(\ref{Lang1v}), it is assumed that $D>2/3$ and $\epsilon>\epsilon_{\mathrm{c}}=3D$~\cite{pikovsky:2002}. Under this assumption, it is easy to show that the parameters $a$ and $b$ are strictly positive.

The Langevin equation defined by Eq.~(\ref{Lang1v}) is formally identical to that of an overdamped Brownian particle moving in a time-periodic potential of period $T$, with a noise term of strength $D/N$ which depends on the system size
$N$. The potential periodically switches between two values, $\smash{U_{-}(x)=b X^4/4-a X^2/2-A X}$ and $\smash{U_{+}(x)=b X^4/4-a X^2/2+A X}$. 
The value $U_{-}(x)$ corresponds to the first half of a period, whereas the value $U_{+}(x)$ corresponds to the second one. 

In the absence of external driving, i.e., for $A=0$,  $U_{\pm}(x)$
is bistable, as $a$ and $b$ are strictly positive. For $A\neq 0$, the potentials $U_{\pm}(x)$ are bistable as long as $\smash{A<A_{\mathrm{th}}=2\sqrt{a^3/(27 b)}}$ (subthreshold external drivings). In this case, the potential $U_{-}(x)$ [respectively, $U_{+}(x)$] possesses two minima at $q_{-1}<0$ and $q_{+1}>0$ (respectively, at $-q_{+1}<0$ and $-q_{-1}>0$), and a maximum at $q_0$ (respectively, at $-q_0$).  The locations $q_{\ell}$ can be calculated using the expression $\smash{q_{\ell}=2\sqrt{a/(3b)}\cos\left\{\left[\nu+2\pi(\ell+2)\right]/3\right\}}$, with $\nu=\arccos(A/A_{\mathrm{th}})$ and $\ell=-1$, $0$, and $+1$. Henceforth, we will only consider subthreshold external drivings. Applying Kramers' rate theory~\cite{hanggi:1990}, the rates of escape $\gamma_{\pm}$ can be expressed as $\smash{\gamma_{\pm}=\omega_{\mathrm{M}}\omega_{\pm}e^{-N E_{\pm}/D}/(2\pi)}$, where $\smash{\omega_{\mathrm{M}}=(a-3 b q_{0}^2)^{1/2}}$, $\smash{\omega_{\pm}=(3 b q_{\pm 1}^2-a})^{1/2}$, and $E_{\pm}=U_{-}(q_0)-U_{-}(q_{\pm 1})$~\cite{casado:2003}.

The effective Langevin equation in Eq.~(\ref{Lang1v}) can be used  to identify ranges of parameter values for which system size synchronization might be expected. To this end, it is first necessary to express Eq.~(\ref{Lang1v}) in a more convenient form  by introducing the rescaled variables $\smash{\tilde{X}=\sqrt{b/a}}X$ and $\smash{\tilde{t}=a t}$.  It can easily be seen that the rescaled stochastic process $\smash{\tilde{X}(\tilde{t})}$ satisfies a Langevin equation analogous to the one considered in Ref.~\cite{casado1:2005}, but for the rescaled values of the amplitude $\smash{\tilde{A}=A\sqrt{b/a^3}}$, noise strength $\smash{\tilde{D}=D b/(N a^2)}$, and frequency $\smash{\tilde{\omega}=\omega/a}$. For given values of the parameters  $\epsilon$ and $N$, these expressions permit to establish a mapping between the parameters $\{\tilde{A},\tilde{D},\tilde{\omega}\}$ considered in Ref.~\cite{casado1:2005} and the parameters $\{A,D,\omega\}$ used in the present work. With this mapping, we can estimate values of $\{A,D,\omega\}$ for which system size synchronization might be expected from the knowledge of values for $\{\tilde{A},\tilde{D},\tilde{\omega}\}$ for which stochastic synchronization has been observed, e.g.,  the set of values used in Ref.~\cite{casado1:2005}. The parameter values in Figs.~\ref{Fig1}, \ref{Fig2}, \ref{Fig3}, and \ref{Fig4} have been determined using this procedure.

\section{Numerical results}
We have carried out numerical simulations of the $N$ coupled Langevin equations given in Eq.~({\ref{Eq1}), as well as of the effective Langevin equation for the single collective variable in Eq.~({\ref{Lang1v}). For the first case, we generate a large number $M$ of random realizations for the $N$ variables, all of them starting from the same initial condition $x_j(0)=X_{\mathrm{th}}=0.6$, with $j=1,2,\ldots, N $. For each of those realizations, the global variable $X(t)$ is evaluated. We have checked numerically that $X(t)$ shows a bistable random  behavior. Panel~(a) in Fig.~\ref{Fig2} shows this bistability. For each trajectory $\alpha=1,\dots,M$ the filtering ideas described above allows us to construct the filtered process $\chi_{\alpha}(t)$ [see panel~(b) in Fig.~\ref{Fig2}], the number of jumps  $\mathcal{N}_{\alpha}(t)$ within the interval $(0,t]$, and the phase $\varphi_{\alpha}(t)=\pi \mathcal{N}_{\alpha}(t)$. Averaging over the $M$ realizations and using Eqs.~(\ref{freq}) and (\ref{Diffcoeff}), we can estimate the output frequency and phase diffusion coefficient. For the case of the Langevin equation for the collective variable in Eq.~({\ref{Lang1v}), the numerical procedure is analogous to the one just described except for the fact that the simulation directly provides the collective variable information.

			\begin{figure}[t]
				\includegraphics[scale=.45]{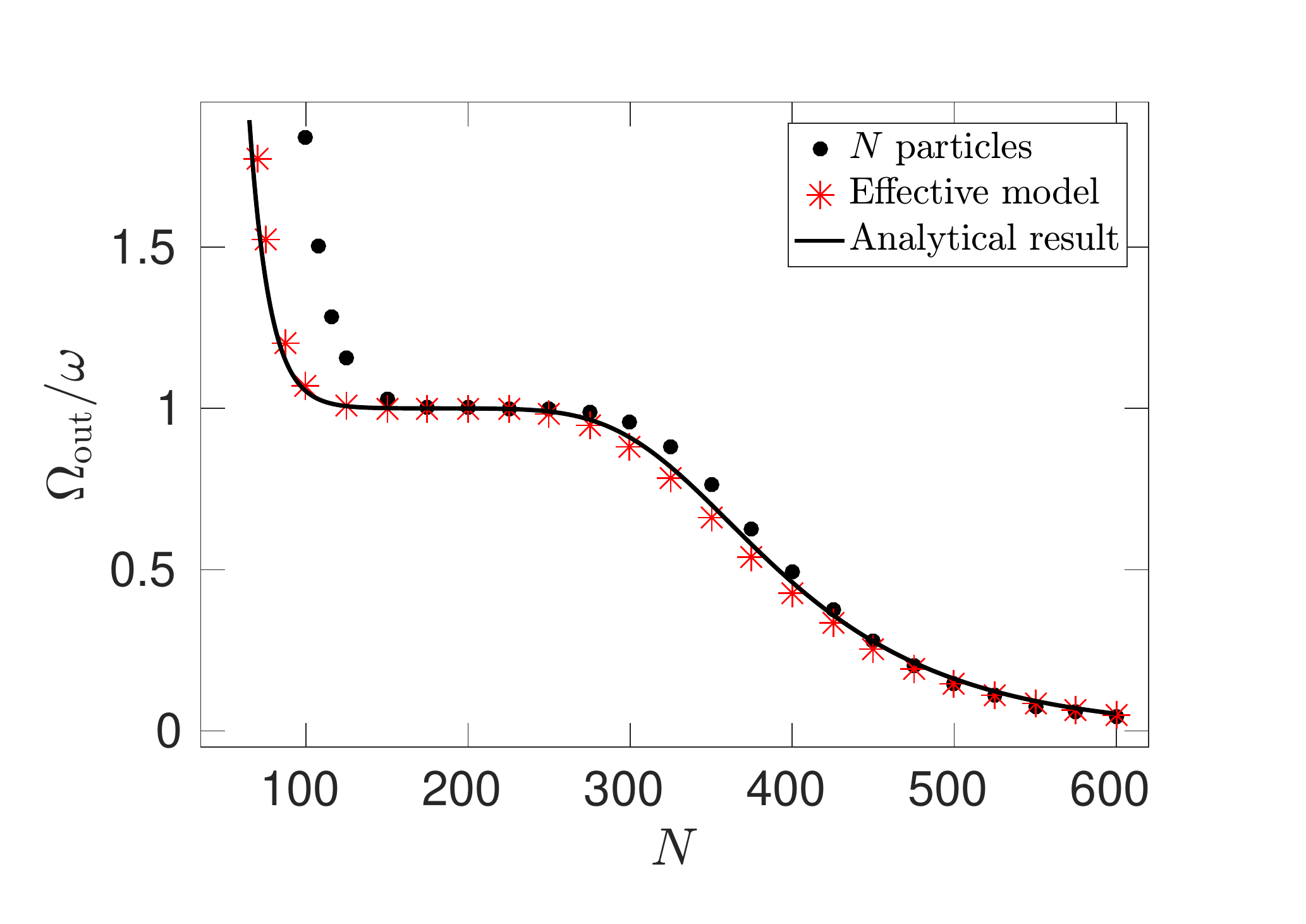}
				\caption{Plot of the ratio of the averaged output frequency, $\Omega_{\mathrm{out}}$, and the driving one, $\omega$, vs. the system size $N$ for $A=0.03$, $D=0.7$, $\omega=0.001$, and $\epsilon=2.7$. Filled circles indicate the results obtained from the simulation of the $N$ coupled Langevin equations in Eq.~(\ref{Eq1}). Stars correspond to the results obtained from the simulation of the effective Langevin equation in Eq.~(\ref{Lang1v}). The solid line describes the analytical result obtained from Eq.~(\ref{anal_freq}).}
				\label{Fig3}
			\end{figure}	
			
			Figure~\ref{Fig3} illustrates what we have called system size frequency locking, i.e., the existence of a range of $N$ values for
			which the output frequency matches the input one. This is one of the characteristics of system size synchronization. With the filled circles, and for the parameter values indicated in the figure caption, we depict the results obtained with the numerical simulations of the whole set of equations in Eq.~(\ref{Eq1}). The stars correspond to the numerical simulation of the Langevin equation for the global variable $X(t)$ in Eq.~(\ref{Lang1v}). The solid line represents the analytical result in Eq.~(\ref{anal_freq}).

			It is interesting to observe that
			the effective Langevin equation for the global variable in Eq.~(\ref{Lang1v}) is able to describe, at least qualitatively, the system size frequency locking. The analytical and numerical simulation results based on the effective Langevin equation agree quite well. The quantitative disagreement between the approximate Langevin
			equation results and those provided by the simulations of the whole set of equations is
			to be expected for small values of $N$. This is so, as the approximations leading
			to the effective Langevin equation are assumed to be  
			valid for large systems. As the system size increases the simulations and the analytical results agree very well.

			The logarithm of the averaged phase diffusion coefficient is presented in  Fig.~\ref{Fig4}. Again, the filled circles correspond to the results obtained with the numerical simulations of the whole system in Eq.~(\ref{Eq1}), the stars to the numerical simulation of the effective Langevin equation in Eq.~(\ref{Lang1v}), and the solid line depicts the analytical result in Eq.~(\ref{anal_diff}). Within the range of $N$ values where the output and driving frequency are locked, we also observe a system size induced phase locking characterized by the fact that the phase diffusion coefficient reaches a rather deep minimum. This feature is another characteristic of system size synchronization. The observation that the dispersion of the jump process, gauged by the phase diffusion coefficient, is so small, is a good indication of a very solid system size synchronization. The discrepancies for small values of $N$ between the many particle simulations and the results based on the effective Langevin equation in Eq.~(\ref{Lang1v}) are larger than for the output frequency shown in Fig.~\ref{Fig3}. This can be understood on the basis that the phase diffusion coefficient involves a second order moment of the jump process, whereas the output frequency only requires the evaluation of the first moment. In any case, both the numerical and analytical treatments of the effective Langevin equation provide a rather good qualitative description of the phenomenology.

			\begin{figure}[t]
				\centering{\includegraphics[scale=.45]{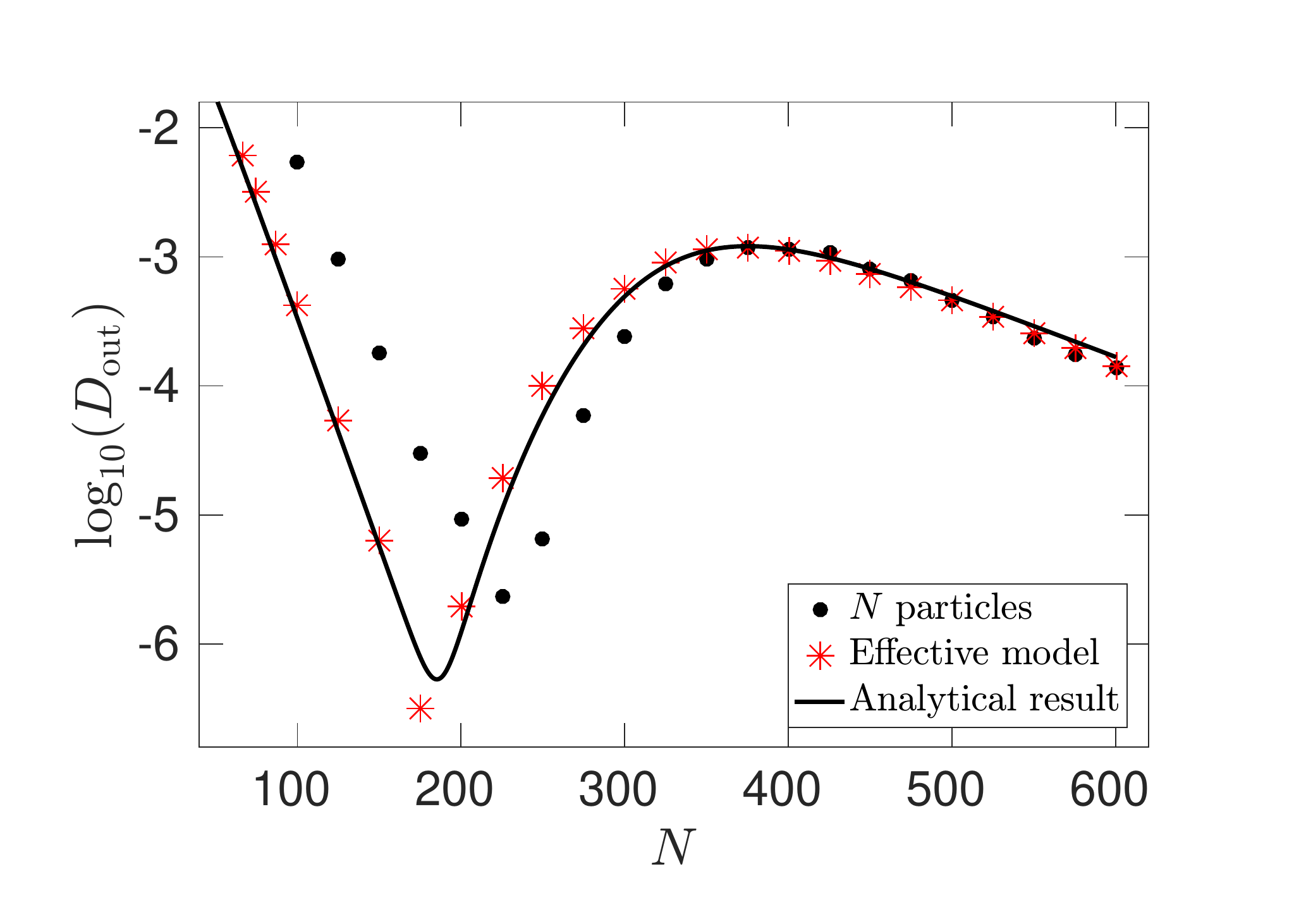}}
				\caption{Plot of the logarithm of the averaged phase diffusion coefficient, $\mathrm{log}_{10}(D_{\mathrm{out}})$, vs. the system size $N$ for $A=0.03$, $D=0.7$, $\omega=0.001$, and $\epsilon=2.7$. Filled circles indicate the results obtained from the simulation of the $N$ coupled Langevin equations in Eq.~(\ref{Eq1}). Stars correspond to the results obtained from the simulation of the effective Langevin equation in Eq.~(\ref{Lang1v}). The solid line describes the analytical result obtained from Eq.~(\ref{anal_diff}).}
				\label{Fig4}
			\end{figure}
		
		\section{Conclusions}
		In this work, we have brought out the existence of a type of synchronization, which we
		have termed \textit{system size synchronization}, for a set of
		coupled noisy elements driven by an external time-periodic force. This phenomenon is quantified in terms of an averaged output frequency and an averaged phase diffusion coefficient. Within an adequate range of $N$ values, the output and the external driving frequencies are locked, and, simultaneously, the phase diffusion coefficient has a very deep minimum.  
		
		Our analytical approximation leads us to conjecture that, in general, for complex stochastic systems others than the one considered here, two main ingredients for the observation of system size synchronization are needed.  A first ingredient is the possibility of defining a global variable showing a dichotomic behavior. A second ingredient is that the transitions between its two values are governed by rates involving the combined action of a size scaled effective noise and the weak applied force.
		
		We think that the present work opens a new perspective on the topic of synchronization which might be of interest in different areas of physical, biological, or medical sciences.
\begin{acknowledgments}
	J.C.-P. acknowledges financial support from the Ministerio de Econom\'{\i}a y Competitividad of Spain through Project No. FIS2017-86478-P. J.C.-P. and M.M acknowledge support from the Junta de Andaluc\'{\i}a.
\end{acknowledgments}

%

\end{document}